\documentclass
[twocolumn,byrevtex,showpacs,showkeys,pra,superscriptaddress]{revtex4}%
\pdfoutput=1
\usepackage{amsfonts}
\usepackage{amsmath}
\usepackage{amssymb}
\usepackage{graphicx}%
\providecommand{\U}[1]{\protect\rule{.1in}{.1in}}
\newtheorem{theorem}{Theorem}

\newenvironment{proof}[1][Proof]{\noindent\textbf{#1.} }{\ \rule{0.5em}{0.5em}}
\begin{document}
\preprint{ }
\title[ ]{Extra Shared Entanglement Reduces Memory Demand in Quantum Convolutional Coding}
\author{Mark M. Wilde}
\affiliation{Centre for Quantum Technologies, National University of Singapore, 3 Science
Drive 2, Singapore 117543}
\affiliation{Center for Quantum Information Science and Technology, Communication Sciences
Institute, Department of Electrical Engineering, University of Southern
California, Los Angeles, California, USA 90089}
\author{Todd A. Brun}
\affiliation{Center for Quantum Information Science and Technology, Communication Sciences
Institute, Department of Electrical Engineering, University of Southern
California, Los Angeles, California, USA 90089}

\begin{abstract}
We show how extra entanglement shared between sender and receiver reduces the
memory requirements for a general entanglement-assisted quantum convolutional
code. We construct quantum convolutional codes with good error-correcting
properties by exploiting the error-correcting properties of an arbitrary basic
set of Pauli generators. The main benefit of this particular construction is
that there is no need to increase the frame size of the code when extra shared
entanglement is available. Then there is no need to increase the memory
requirements or circuit complexity of the code because the frame size of the
code is directly related to these two code properties. Another benefit,
similar to results of previous work in entanglement-assisted convolutional
coding, is that we can import an arbitrary classical quaternary code for use
as an entanglement-assisted quantum convolutional code. The rate and
error-correcting properties of the imported classical code translate to the
quantum code. We provide an example that illustrates how to import a classical
quaternary code for use as an entanglement-assisted quantum convolutional
code. We finally show how to \textquotedblleft piggyback\textquotedblright%
\ classical information to make use of the extra shared entanglement in the code.

\end{abstract}
\volumeyear{2008}
\volumenumber{ }
\issuenumber{ }
\eid{ }
\date{\today}
\startpage{1}
\endpage{10}
\maketitle

\section{Introduction}

Quantum convolutional coding is a useful technique for encoding a stream of
quantum information before sending it over a noisy quantum communication
channel
\cite{PhysRevLett.91.177902,arxiv2004olliv,isit2006grassl,ieee2006grassl,ieee2007grassl,isit2005forney,ieee2007forney,cwit2007aly,arx2007aly}%
. Ollivier and Tillich started the investigations into the theory of quantum
convolutional coding \cite{PhysRevLett.91.177902,arxiv2004olliv}\ and many
authors have since contributed to a general theory of quantum convolutional
coding
\cite{isit2006grassl,ieee2006grassl,ieee2007grassl,isit2005forney,ieee2007forney,cwit2007aly,arx2007aly}%
.

Quantum convolutional codes have several key benefits. The
performance/complexity trade-off for a quantum convolutional code is superior
to that of a block code that encodes the same number of information qubits
\cite{ieee2007forney}. Some benefits of their classical counterparts
\cite{book1999conv} carry over to quantum convolutional codes---e.g., the
encoding circuits for both quantum and classical convolutional codes have a
periodic structure.

Entanglement-assisted quantum error-correcting codes exploit entanglement
shared between sender and receiver \cite{science2006brun}. The
entanglement-assisted communication paradigm assumes that a sender and
receiver share a set of noiseless ebits, the sender can use her half of these
ebits in the encoding procedure, and the receiver's half of the shared ebits
remain noiseless throughout communication. The benefit of
entanglement-assisted stabilizer codes is that the sender and receiver can
exploit the error-correcting properties of an arbitrary set of Pauli
generators rather than restrict themselves to a commuting set. Another benefit
is that entanglement boosts the quantum communication rate of the
entanglement-assisted quantum code constructed from either two classical
binary codes or one classical quaternary code.

We have recently constructed entanglement-assisted quantum convolutional codes
that admit a Calderbank-Shor-Steane (CSS) structure \cite{arx2007wildeEAQCC}
and others that admit a more general structure \cite{arx2008wildeGEAQCC}.
These papers develop several ways of encoding and decoding
entanglement-assisted quantum convolutional codes with operations that are
both finite depth and infinite depth. The upshot of these constructions is
that we can import two arbitrary classical binary convolutional codes for use
as a CSS\ entanglement-assisted quantum convolutional code
\cite{arx2007wildeEAQCC}\ or we can import an arbitrary classical quaternary
convolutional code for use as an entanglement-assisted quantum convolutional
code \cite{arx2008wildeGEAQCC}. The rates and error-correcting properties of
the classical codes translate directly to the quantum code.

In Ref.~\cite{arx2008wildeGEAQCC}, we showed how to accomodate entangled bits
in an entanglement-assisted quantum convolutional code by increasing the
code's frame size. (A quantum convolutional code partitions quantum data into
uniformly-sized \textit{frames} consisting of ebits, ancilla qubits, and
information qubits that a periodic encoding circuit acts on.) Increasing the frame size of
the code is undesirable because it increases the complexity of the encoding and
decoding circuits and demands a larger memory for the code.

In this paper, we show that there
is no need to increase the frame size of the entanglement-assisted quantum
convolutional code if one makes use of extra shared entanglement.
The result in Ref.~\cite{arx2008wildeGEAQCC} and
our findings in this paper seem to point to a trade-off between efficient use of
entanglement and that of circuit resources for non-CSS\ entanglement-assisted
quantum convolutional codes. We again
focus on more general entanglement-assisted quantum convolutional codes that
do not have the CSS\ structure. The outcome of this work is that we can import
an arbitrary classical quaternary convolutional code for use as an
entanglement-assisted quantum convolutional code with the benefit that it is
not necessary to increase the frame size of the code. The drawback of the
technique in this paper is that it does not make the most efficient use of entanglement.

\section{Main Theorem}

The proof of our main theorem below outlines how to encode a stream of
information qubits, ancilla qubits, and shared ebits so that the encoded
qubits have the error-correcting properties of an arbitrary set of Paulis. The
receiver may employ an error estimation algorithm such as Viterbi decoding
\cite{itit1967viterbi}\ to determine the most likely errors that the noisy
quantum communication channel induces on the encoded stream. We then show how
to decode the encoded qubit stream so that the information qubits become
available at the receiving end of the channel.

The encoding circuits in the proof of our theorem employ both finite-depth and
infinite-depth operations. The decoding circuits employ finite-depth
operations only. Finite-depth operations take a finite-weight stabilizer to
one with finite weight. Infinite-depth operations take \textit{some}
finite-weight stabilizer generators to ones with infinite weight.

Infinite-depth operations can lead to catastrophic error propagation
\cite{PhysRevLett.91.177902,arxiv2004olliv,isit2006grassl,arx2007wildeEAQCC}.
In our proof below, we restrict infinite-depth operations to act on qubits
before sending them over the noisy channel. Catastrophic error propagation
does not occur under the ideal circumstance when the operations in the
encoding circuit are noiseless.

Theorem~\ref{thm:main} below begins with a \textquotedblleft quantum check
matrix\textquotedblright\ that consists of a set of Pauli sequences with
desirable error-correcting properties \cite{arx2007wildeEAQCC}. This quantum
check matrix represents an arbitrary basic set of Pauli generators and thus
does not necessarily correspond to a commuting stabilizer. The proof of the
theorem shows how to incorporate ebits so that the sender realizes the same
quantum check matrix for her qubits and the sender and receiver's set of
generators form a valid commuting stabilizer.

The entries of the quantum check matrix in Theorem~\ref{thm:main} are binary
polynomials. The usual way to represent a quantum code is with a tensor
product of Pauli matrices, but it is more convenient for our purposes to
perform all manipulations with matrices of binary polynomials rather than with
tensor products of Pauli matrices (see Refs.
\cite{arxiv2004olliv,ieee2006grassl,isit2006grassl,ieee2007forney,arx2007wilde,arx2007wildeEAQCC}
for details of this isomorphism). Consider the quantum check matrix in
Theorem~\ref{thm:main}. The matrix on the left (right)\ side of the vertical
bar includes binary polynomials corresponding to $Z$ ($X$) Pauli operators.
Also, the delay operator $D$ gives a simple way of obtaining all the
generators that the quantum check matrix represents. Multiplication of any of
the rows in the matrix by any power of $D$ gives a corresponding generator.

We perform two types of manipulations on the below quantum check matrix: row
operations and column operations. Row operations do not change the
error-correcting properties of the code. Column operations correspond to
quantum circuit elements in the shift-invariant Clifford group
\cite{ieee2006grassl,isit2006grassl}\ and do not change the shifted symplectic
product \cite{arx2007wilde,arx2007wildeEAQCC}\ between the rows of the quantum
check matrix. Both row and column operations are crucial for arriving at the
proper decomposition of the below quantum check matrix.

\begin{theorem}
\label{thm:main}Suppose we would like to exploit the error-correcting
properties of the generators in the following quantum check matrix:%
\[
S\left(  D\right)  =\left[  \left.
\begin{array}
[c]{c}%
Z\left(  D\right)
\end{array}
\right\vert
\begin{array}
[c]{c}%
X\left(  D\right)
\end{array}
\right]  \in\mathbb{F}_{2}\left[  D\right]  ^{\left(  n-k\right)  \times2n},
\]
where $S\left(  D\right)  $ is of full rank and does not necessarily form a
commuting stabilizer. Then an $\left[  \left[  n,k;c\right]  \right]  $
entanglement-assisted quantum convolutional code exists that has the same
error-correcting properties as the above quantum check matrix $S\left(
D\right)  $ where $c=\ $rank$\left(  X\left(  D\right)  \right)  $ (an
$\left[  \left[  n,k;c\right]  \right]  $ code is one that encodes $k$
information qubits per frame into $n$ \textquotedblleft
channel\textquotedblright\ qubits per frame while consuming $c$ ebits per
frame \cite{science2006brun,arx2007wildeEAQCC,arx2008wildeGEAQCC}).
\end{theorem}

\begin{proof}
The first part of the proof in (\ref{eq:first-finite-depth-encode}%
-\ref{eq:big-QCM}) involves decomposing the above check matrix using CNOT and
SWAP\ gates chosen according to the Smith algorithm \cite{book1999conv} (it
employs intermediate Hadamard gates as well). The resulting check matrix in
(\ref{eq:big-QCM})\ is then no longer decomposable using finite-depth
operations only. We then start with a matrix representing unencoded ebits and
show how to perform infinite-depth encoding operations in order to obtain part
of the check matrix in (\ref{eq:big-QCM}). The last part of the proof shows
how to decode the information qubits properly using finite-depth operations
only. The order of the steps in this proof is similar to the order of the
steps in Refs.~\cite{arx2007wildeEAQCC,arx2008wildeGEAQCC}, but the details of
the proof are different. The details of performing column operations with CNOT
and Hadamard gates, and infinite-depth operations corresponding to multiplying
by rational polynomials, are also in those papers.

Suppose the Smith form \cite{book1999conv}\ of $X\left(  D\right)  $ is as
follows:%
\begin{equation}
X\left(  D\right)  =A\left(  D\right)  \left[
\begin{array}
[c]{ccc}%
\Gamma_{1}\left(  D\right)  & 0 & 0\\
0 & \Gamma_{2}\left(  D\right)  & 0\\
0 & 0 & 0
\end{array}
\right]  B\left(  D\right)  , \label{eq:first-finite-depth-encode}%
\end{equation}
where $A\left(  D\right)  $ is $\left(  n-k\right)  \times\left(  n-k\right)
$, $B\left(  D\right)  $ is $n\times n$, $\Gamma_{1}\left(  D\right)  $ is an
$s\times s$ diagonal matrix whose entries are powers of $D$, and $\Gamma
_{2}\left(  D\right)  $ is a $\left(  c-s\right)  \times\left(  c-s\right)  $
diagonal matrix whose entries are arbitrary polynomials. Perform the row
operations in $A^{-1}\left(  D\right)  $ and the column operations in
$B^{-1}\left(  D\right)  $ on $S\left(  D\right)  $. The quantum check matrix
$S\left(  D\right)  $\ becomes%
\begin{equation}
\left[
\begin{array}
[c]{c}%
E\left(  D\right)
\end{array}
\left\vert
\begin{array}
[c]{ccc}%
\Gamma_{1}\left(  D\right)  & 0 & 0\\
0 & \Gamma_{2}\left(  D\right)  & 0\\
0 & 0 & 0
\end{array}
\right.  \right]  ,
\end{equation}
where $E\left(  D\right)  =A^{-1}\left(  D\right)  Z\left(  D\right)
B^{T}\left(  D^{-1}\right)  $. Suppose $E_{1}\left(  D\right)  $ is the first
$c$ columns of $E\left(  D\right)  $ and $E_{2}\left(  D\right)  $ is the next
$n-c$ columns of $E\left(  D\right)  $ so that the quantum check matrix is as
follows:%
\begin{equation}
\left[
\begin{array}
[c]{cc}%
E_{1}\left(  D\right)  & E_{2}\left(  D\right)
\end{array}
\left\vert
\begin{array}
[c]{ccc}%
\Gamma_{1}\left(  D\right)  & 0 & 0\\
0 & \Gamma_{2}\left(  D\right)  & 0\\
0 & 0 & 0
\end{array}
\right.  \right]  .
\end{equation}
Perform Hadamard gates on the last $n-c$ qubits so that the quantum check
matrix becomes%
\begin{equation}
\left[
\begin{array}
[c]{cc}%
E_{1}\left(  D\right)  & 0
\end{array}
\left\vert
\begin{array}
[c]{ccc}%
\Gamma_{1}\left(  D\right)  & 0 & E_{2,1}\left(  D\right) \\
0 & \Gamma_{2}\left(  D\right)  & E_{2,2}\left(  D\right) \\
0 & 0 & E_{2,3}\left(  D\right)
\end{array}
\right.  \right]  ,
\end{equation}
where%
\begin{equation}
E_{2}\left(  D\right)  =\left[
\begin{array}
[c]{c}%
E_{2,1}\left(  D\right) \\
E_{2,2}\left(  D\right) \\
E_{2,3}\left(  D\right)
\end{array}
\right]  .
\end{equation}
Perform CNOT\ operations from the first $s$ qubits to the last $n-c$ qubits to
clear the entries in $E_{2,1}\left(  D\right)  $. The quantum check matrix
becomes%
\begin{equation}
\left[
\begin{array}
[c]{cc}%
E_{1}\left(  D\right)  & 0
\end{array}
\left\vert
\begin{array}
[c]{ccc}%
\Gamma_{1}\left(  D\right)  & 0 & 0\\
0 & \Gamma_{2}\left(  D\right)  & E_{2,2}\left(  D\right) \\
0 & 0 & E_{2,3}\left(  D\right)
\end{array}
\right.  \right]  .
\end{equation}
The Smith form of $E_{2,3}\left(  D\right)  $ is as follows:%
\begin{equation}
E_{2,3}\left(  D\right)  =A_{E}\left(  D\right)  \left[
\begin{array}
[c]{cc}%
\Gamma\left(  D\right)  & 0
\end{array}
\right]  B_{E}\left(  D\right)  ,
\end{equation}
where $A_{E}\left(  D\right)  $ is $\left(  n-k-c\right)  \times\left(
n-k-c\right)  $, $B_{E}\left(  D\right)  $ is $\left(  n-c\right)
\times\left(  n-c\right)  $, and $\Gamma\left(  D\right)  $ is a $\left(
n-k-c\right)  \times\left(  n-k-c\right)  $\ diagonal matrix whose entries are
polynomials. The Smith form of $E_{2,3}\left(  D\right)  $\ is full rank
because\ the original quantum check matrix $S\left(  D\right)  $\ is full
rank. Perform the row operations in $A_{E}^{-1}\left(  D\right)  $ and the
column operations in $B_{E}^{-1}\left(  D\right)  $. The quantum check matrix
becomes%
\begin{equation}
\left[
\begin{array}
[c]{cc}%
E_{1}^{\prime}\left(  D\right)  & 0
\end{array}
\left\vert
\begin{array}
[c]{cccc}%
\Gamma_{1}\left(  D\right)  & 0 & 0 & 0\\
0 & \Gamma_{2}\left(  D\right)  & E_{2,2a}^{^{\prime}}\left(  D\right)  &
E_{2,2b}^{^{\prime}}\left(  D\right) \\
0 & 0 & \Gamma\left(  D\right)  & 0
\end{array}
\right.  \right]  ,
\end{equation}
where%
\begin{align}
E_{1}^{\prime}\left(  D\right)   &  =\left[
\begin{array}
[c]{cc}%
I & 0\\
0 & A_{E}^{-1}\left(  D\right)
\end{array}
\right]  E_{1}\left(  D\right)  ,\\
E_{2,2}^{^{\prime}}\left(  D\right)   &  =E_{2,2}\left(  D\right)  B_{E}%
^{-1}\left(  D\right) \\
&  =\left[
\begin{array}
[c]{cc}%
E_{2,2a}^{^{\prime}}\left(  D\right)  & E_{2,2b}^{^{\prime}}\left(  D\right)
\end{array}
\right]  .
\end{align}
Perform a modified version of the Smith algorithm to reduce the $\left(
c-s\right)  \times\left(  n-c\right)  $ matrix $E_{2,2b}^{^{\prime}}\left(
D\right)  $ to a lower triangular form \cite{arx2007wildeEAQCC}. This modified
algorithm uses only column operations to transform%
\begin{equation}
E_{2,2b}^{^{\prime}}\left(  D\right)  \rightarrow\left[
\begin{array}
[c]{cc}%
L\left(  D\right)  & 0
\end{array}
\right]  .
\end{equation}
where $L\left(  D\right)  $ is $\left(  c-s\right)  \times\left(  c-s\right)
$ and the null matrix is $\left(  c-s\right)  \times\left(  n+s-2c\right)  $.
The quantum check matrix becomes%
\begin{equation}
\left[
\begin{array}
[c]{cc}%
E_{1}^{\prime}\left(  D\right)  & 0
\end{array}
\left\vert
\begin{array}
[c]{ccccc}%
\Gamma_{1}\left(  D\right)  & 0 & 0 & 0 & 0\\
0 & \Gamma_{2}\left(  D\right)  & E_{2,2a}^{^{\prime}}\left(  D\right)  &
L\left(  D\right)  & 0\\
0 & 0 & \Gamma\left(  D\right)  & 0 & 0
\end{array}
\right.  \right]  . \label{eq:big-QCM}%
\end{equation}
We have now completed the decomposition of the quantum check matrix with
column and row operations.

We turn to showing how to encode a certain quantum check matrix that proves to
be useful in encoding the above quantum check matrix. Consider the following
quantum check matrix:%
\begin{equation}
\left[
\begin{array}
[c]{cc}%
I & 0\\
0 & 0
\end{array}
\left\vert
\begin{array}
[c]{cc}%
0 & 0\\
\Gamma_{2}\left(  D\right)  & L\left(  D\right)
\end{array}
\right.  \right]  , \label{eq:desired-QCM}%
\end{equation}
where $\Gamma_{2}\left(  D\right)  $ and $L\left(  D\right)  $ are from the
matrix in (\ref{eq:big-QCM})\ and each of them, the identity matrix, and the
null matrices have dimension $\left(  c-s\right)  \times\left(  c-s\right)  $.

We use a method for encoding the quantum check matrix in (\ref{eq:desired-QCM}%
)\ similar to the method outlined in Ref.~\cite{arx2007wildeEAQCC}\ for the
second class of CSS\ entanglement-assisted quantum convolutional codes. We
begin with a set of $c-s$ ebits and $c-s$ information qubits. The following
matrix stabilizes the ebits:%
\begin{equation}
\left[  \left.
\begin{array}
[c]{ccc}%
I & I & 0\\
0 & 0 & 0
\end{array}
\right\vert
\begin{array}
[c]{ccc}%
0 & 0 & 0\\
I & I & 0
\end{array}
\right]  , \label{eq:first-inf-depth-encode}%
\end{equation}
where Bob has the $c-s$ qubits in the leftmost column of each submatrix, Alice
has the $2(c-s)$ qubits in the right two columns, and each block is
$(c-s)\times(c-s)$. The following matrix represents the logical operators for
the information qubits, and gives a useful way of tracking the information
qubits while processing them:%
\begin{equation}
\left[  \left.
\begin{array}
[c]{ccc}%
0 & 0 & I\\
0 & 0 & 0
\end{array}
\right\vert
\begin{array}
[c]{ccc}%
0 & 0 & 0\\
0 & 0 & I
\end{array}
\right]  . \label{eq:first-inf-depth-encode-info}%
\end{equation}
Tracking the information-qubit matrix helps to confirm that the information
qubits decode properly at the receiver's end \cite{arx2007wildeEAQCC}.

We now track both the above stabilizer and the information-qubit matrix as
they progress through some encoding operations. Alice performs CNOT\ gates
from her first $c-s$ qubits to her next $c-s$ qubits. These gates multiply the
middle $c-s$ columns of the \textquotedblleft X\textquotedblright\ matrix by
$L\left(  D\right)  $ and add the result to the last $c-s$ columns, and
multiply the last $c-s$ columns of the \textquotedblleft Z\textquotedblright%
\ matrix by $L^{T}\left(  D^{-1}\right)  $ and add the result to the last
$c-s$ columns. The stabilizer becomes%
\begin{equation}
\left[  \left.
\begin{array}
[c]{ccc}%
I & I & 0\\
0 & 0 & 0
\end{array}
\right\vert
\begin{array}
[c]{ccc}%
0 & 0 & 0\\
I & I & L\left(  D\right)
\end{array}
\right]  ,
\end{equation}
and the information-qubit matrix becomes%
\begin{equation}
\left[  \left.
\begin{array}
[c]{ccc}%
0 & L^{T}\left(  D^{-1}\right)  & I\\
0 & 0 & 0
\end{array}
\right\vert
\begin{array}
[c]{ccc}%
0 & 0 & 0\\
0 & 0 & I
\end{array}
\right]  .
\end{equation}
Alice performs infinite-depth operations on her first $c-s$ qubits
corresponding to the rational polynomials $\gamma_{2,1}^{-1}\left(
D^{-1}\right)  $, $\ldots$, $\gamma_{2,c-s}^{-1}\left(  D^{-1}\right)  $ in
$\Gamma_{2}^{-1}\left(  D^{-1}\right)  $. These operations multiply the middle
$c-s$ columns of the \textquotedblleft Z\textquotedblright\ matrix by
$\Gamma_{2}^{-1}\left(  D^{-1}\right)  $ and multiply the middle $c-s$ columns
of the \textquotedblleft X\textquotedblright\ matrix by $\Gamma_{2}\left(
D\right)  $. The stabilizer matrix becomes%
\begin{equation}
\left[  \left.
\begin{array}
[c]{ccc}%
I & \Gamma_{2}^{-1}\left(  D^{-1}\right)  & 0\\
0 & 0 & 0
\end{array}
\right\vert
\begin{array}
[c]{ccc}%
0 & 0 & 0\\
I & \Gamma_{2}\left(  D\right)  & L\left(  D\right)
\end{array}
\right]  , \label{eq:last-inf-depth-encode-stab}%
\end{equation}
and the information-qubit matrix becomes%
\begin{equation}
\left[  \left.
\begin{array}
[c]{ccc}%
0 & L^{T}\left(  D^{-1}\right)  \Gamma_{2}^{-1}\left(  D^{-1}\right)  & I\\
0 & 0 & 0
\end{array}
\right\vert
\begin{array}
[c]{ccc}%
0 & 0 & 0\\
0 & 0 & I
\end{array}
\right]  . \label{eq:last-inf-depth-encode}%
\end{equation}
Alice's part of the above stabilizer matrix is equivalent to the quantum check
matrix in (\ref{eq:desired-QCM}) by row operations (premultiplying the first
set of rows by $\Gamma_{2}\left(  D^{-1}\right)  $.)

We now show how to encode the quantum check matrix in\ (\ref{eq:big-QCM})
using ebits, ancilla qubits, and information qubits. We employ the encoding
technique for the submatrix listed above in (\ref{eq:first-inf-depth-encode}%
-\ref{eq:last-inf-depth-encode}) and use some other techniques as well.
Suppose that we have the following matrix that stabilizes a set of $c$ ebits
per frame, $n-k-c$ ancilla qubits per frame, and $k$ information qubits per
frame:%
\begin{equation}
\left[
\begin{array}
[c]{ccccccc}%
I & 0 & I & 0 & 0 & 0 & 0\\
0 & I & 0 & I & 0 & 0 & 0\\
0 & 0 & 0 & 0 & 0 & 0 & 0\\
0 & 0 & 0 & 0 & 0 & 0 & 0\\
0 & 0 & 0 & 0 & 0 & 0 & 0
\end{array}
\left\vert
\begin{array}
[c]{ccccccc}%
0 & 0 & 0 & 0 & 0 & 0 & 0\\
0 & 0 & 0 & 0 & 0 & 0 & 0\\
I & 0 & I & 0 & 0 & 0 & 0\\
0 & I & 0 & I & 0 & 0 & 0\\
0 & 0 & 0 & 0 & I & 0 & 0
\end{array}
\right.  \right]  .
\end{equation}
The first and third sets of rows have $s$ rows and correspond to $s$ ebits per
frame, the second and fourth sets of rows have $c-s$ rows and correspond to
$c-s$ ebits per frame, and the last set of $n-k-c$ rows corresponds to $n-k-c$
ancilla qubits per frame. The above matrix has $n+c$ columns on both the
\textquotedblleft Z\textquotedblright\ and \textquotedblleft
X\textquotedblright\ side so it stabilizes $k$ information qubits per frame.
Bob possesses the first $c$ qubits and Alice possesses the next $n$ qubits.
Alice performs the encoding operations in (\ref{eq:first-inf-depth-encode}%
-\ref{eq:last-inf-depth-encode}) to get the following stabilizer:%
\begin{equation}
\left[
\begin{array}
[c]{ccccccc}%
I & 0 & I & 0 & 0 & 0 & 0\\
0 & I & 0 & G\left(  D\right)  & 0 & 0 & 0\\
0 & 0 & 0 & 0 & 0 & 0 & 0\\
0 & 0 & 0 & 0 & 0 & 0 & 0\\
0 & 0 & 0 & 0 & 0 & 0 & 0
\end{array}
\left\vert
\begin{array}
[c]{ccccccc}%
0 & 0 & 0 & 0 & 0 & 0 & 0\\
0 & 0 & 0 & 0 & 0 & 0 & 0\\
I & 0 & I & 0 & 0 & 0 & 0\\
0 & I & 0 & \Gamma_{2}\left(  D\right)  & 0 & L\left(  D\right)  & 0\\
0 & 0 & 0 & 0 & I & 0 & 0
\end{array}
\right.  \right]  , \label{eq:start-decoding}%
\end{equation}
where $G\left(  D\right)  \equiv\Gamma_{2}^{-1}\left(  D^{-1}\right)  $.

We perform several row operations to get the quantum check matrix
in\ (\ref{eq:big-QCM}). Premultiply the middle set of rows by $\Gamma
_{1}\left(  D\right)  $ (one actually does not have to do this row
operation---the result is a \textquotedblleft subcode\textquotedblright\ of
the original code \cite{isit2006grassl}). Premultiply the last set of rows by
$E_{2,2a}^{^{\prime}}\left(  D\right)  $ and add the result to the set of rows
above the last set. Premultiply the last set of rows by $\Gamma\left(
D\right)  $ (one also does not have to do this row operation and the result is
again a subcode). Finally, premultiply the first two sets of rows by%
\[
E_{1}^{\prime\prime}\left(  D\right)  \equiv E_{1}^{\prime}\left(  D\right)
\left[  I\oplus\Gamma_{2}\left(  D^{-1}\right)  \right]  ,
\]
and add the result to the last three sets of rows. The \textquotedblleft
Z\textquotedblright\ side of the quantum check matrix becomes%
\begin{equation}
\left[
\begin{array}
[c]{ccccccc}%
I & 0 & I & 0 & 0 & 0 & 0\\
0 & I & 0 & \Gamma_{2}^{-1}\left(  D^{-1}\right)  & 0 & 0 & 0\\
&  &  &  & 0 & 0 & 0\\
E_{1}^{\prime\prime}\left(  D\right)  &  & E_{1}^{\prime}\left(  D\right)  &
& 0 & 0 & 0\\
&  &  &  & 0 & 0 & 0
\end{array}
\right\vert , \label{eq:final-encoded-Z}%
\end{equation}
and the \textquotedblleft X\textquotedblright\ side becomes%
\begin{equation}
\left\vert
\begin{array}
[c]{ccccccc}%
0 & 0 & 0 & 0 & 0 & 0 & 0\\
0 & 0 & 0 & 0 & 0 & 0 & 0\\
\Gamma_{1}\left(  D\right)  & 0 & \Gamma_{1}\left(  D\right)  & 0 & 0 & 0 &
0\\
0 & \Gamma_{2}\left(  D\right)  & 0 & \Gamma_{2}\left(  D\right)  &
E_{2,2a}^{^{\prime}}\left(  D\right)  & L\left(  D\right)  & 0\\
0 & 0 & 0 & 0 & \Gamma\left(  D\right)  & 0 & 0
\end{array}
\right]  . \label{eq:final-encoded-X}%
\end{equation}

The last three rows of Alice's part of the above quantum check matrix are
equivalent to the quantum check matrix in (\ref{eq:big-QCM}). The first two
sets of rows represent extra shared entanglement that Alice and Bob use to
resolve the anticommutativity present in the original set of generators. Alice
performs all finite-depth encoding operations (column operations)\ in
(\ref{eq:first-finite-depth-encode}-\ref{eq:big-QCM}) in reverse order to
obtain the desired quantum check matrix in the statement of the theorem.

Decoding first consists of performing all the operations in
(\ref{eq:first-finite-depth-encode}-\ref{eq:big-QCM}). We now illustrate a way
to decode the stabilizer in\ (\ref{eq:last-inf-depth-encode-stab}) and
information-qubit matrix in (\ref{eq:last-inf-depth-encode})\ so that the
information qubits appear at the output of the decoding circuit. Bob performs
CNOT\ gates from the first set of qubits to the third set of qubits
corresponding to the entries in $L\left(  D\right)  $. The stabilizer becomes%
\begin{equation}
\left[  \left.
\begin{array}
[c]{ccc}%
I & \Gamma_{2}^{-1}\left(  D^{-1}\right)  & 0\\
0 & 0 & 0
\end{array}
\right\vert
\begin{array}
[c]{ccc}%
0 & 0 & 0\\
I & \Gamma_{2}\left(  D\right)  & 0
\end{array}
\right]  , \label{eq:first-inf-depth-decode}%
\end{equation}
and the information-qubit matrix becomes%
\begin{equation}
\left[  \left.
\begin{array}
[c]{ccc}%
L^{T}\left(  D^{-1}\right)  & L^{T}\left(  D^{-1}\right)  \Gamma_{2}%
^{-1}\left(  D^{-1}\right)  & I\\
0 & 0 & 0
\end{array}
\right\vert
\begin{array}
[c]{ccc}%
0 & 0 & 0\\
0 & 0 & I
\end{array}
\right]  .
\end{equation}
Bob finishes decoding at this point because we can equivalently express the
information-qubit matrix as follows:%
\begin{equation}
\left[  \left.
\begin{array}
[c]{ccc}%
0 & 0 & I\\
0 & 0 & 0
\end{array}
\right\vert
\begin{array}
[c]{ccc}%
0 & 0 & 0\\
0 & 0 & I
\end{array}
\right]  , \label{eq:last-inf-depth-decode}%
\end{equation}
by multiplying the first $c-s$ rows of the stabilizer by $L^{T}\left(
D^{-1}\right)  $ and adding to the first $c-s$ rows of the information-qubit
matrix. The information qubits are available at the receiving end of the
channel because the above information-qubit matrix is equivalent to the
original one in (\ref{eq:first-inf-depth-encode-info}).

The above code is an $\left[  \left[  n,k;c\right]  \right]  $
entanglement-assisted code because the code uses a noisy quantum communication
channel $n$ times per frame to send $k$ information qubits per frame. The
parameter $c=\ $rank$\left(  X\left(  D\right)  \right)  $ because the
matrices $\Gamma_{1}\left(  D\right)  $ and $\Gamma_{2}\left(  D\right)  $
determine the rank of the matrix $X\left(  D\right)  $ by the Smith algorithm
\cite{book1999conv}.
\end{proof}

\section{Example}

We present an example of a classical quaternary code over $GF\left(  4\right)
$ that we import for use as an entanglement-assisted quantum convolutional
code \cite{arx2007wilde}:%
\begin{equation}
\left(  \cdots|0000|1\bar{\omega}10|1101|0000|\cdots\right)  .
\end{equation}
The above code is a convolutional version of the classical quaternary block
code from Ref.~\cite{science2006brun}. We multiply the above generator by
$\bar{\omega}$ and $\omega$ as prescribed in Refs.
\cite{ieee1998calderbank,ieee2007forney} and use the following map,%
\begin{equation}
0\rightarrow I,\ \ \omega\rightarrow X,\ \ 1\rightarrow Y,\ \ \bar{\omega
}\rightarrow Z,
\end{equation}
to obtain the following two Pauli generators:%
\begin{align}
&  \left(  \cdots|IIII|ZXZI|ZZIZ|IIII|\cdots\right)  ,\nonumber\\
&  \left(  \cdots|IIII|XYXI|XXIX|IIII|\cdots\right)  .
\end{align}
We write the above two generators as a quantum check matrix:%
\begin{equation}
\left[  \left.
\begin{array}
[c]{cccc}%
1+D & D & 1 & D\\
0 & 1 & 0 & 0
\end{array}
\right\vert
\begin{array}
[c]{cccc}%
0 & 1 & 0 & 0\\
1+D & 1+D & 1 & D
\end{array}
\right]  . \label{eq:multigen-stab}%
\end{equation}

We now show how entanglement helps in implementing the above code as an
entanglement-assisted quantum convolutional code. We encode two qubits per
frame with the help of two ebits. The stabilizer matrix for the unencoded
qubit stream is as follows:%
\begin{equation}
\left[  \left.
\begin{array}
[c]{cccccc}%
0 & 1 & 1 & 0 & 0 & 0\\
1 & 0 & 0 & 1 & 0 & 0\\
0 & 0 & 0 & 0 & 0 & 0\\
0 & 0 & 0 & 0 & 0 & 0
\end{array}
\right\vert
\begin{array}
[c]{cccccc}%
0 & 0 & 0 & 0 & 0 & 0\\
0 & 0 & 0 & 0 & 0 & 0\\
0 & 1 & 1 & 0 & 0 & 0\\
1 & 0 & 0 & 1 & 0 & 0
\end{array}
\right]
\end{equation}
Rows one and three correspond to one ebit and rows two and four correspond to
the other. Multiply row one by $D$ and add the result to row three, multiply
row one by $g\left(  D\right)  =1+D^{-1}+D^{2}$ and add the result to row
four, and multiply row two by $f\left(  D\right)  =1+D^{-2}$ and add the
result to row four. These row operations give the following equivalent
stabilizer:%
\begin{equation}
\left[  \left.
\begin{array}
[c]{cccccc}%
0 & 1 & 1 & 0 & 0 & 0\\
1 & 0 & 0 & 1 & 0 & 0\\
0 & D & D & 0 & 0 & 0\\
f\left(  D\right)  & g\left(  D\right)  & g\left(  D\right)  & f\left(
D\right)  & 0 & 0
\end{array}
\right\vert
\begin{array}
[c]{cccccc}%
0 & 0 & 0 & 0 & 0 & 0\\
0 & 0 & 0 & 0 & 0 & 0\\
0 & 1 & 1 & 0 & 0 & 0\\
1 & 0 & 0 & 1 & 0 & 0
\end{array}
\right]  . \label{eq:unencoded-free-ent-example}%
\end{equation}%
\begin{figure}
[ptb]
\begin{center}
\includegraphics[
natheight=11.846200in,
natwidth=10.133900in,
height=3.7092in,
width=3.3615in
]%
{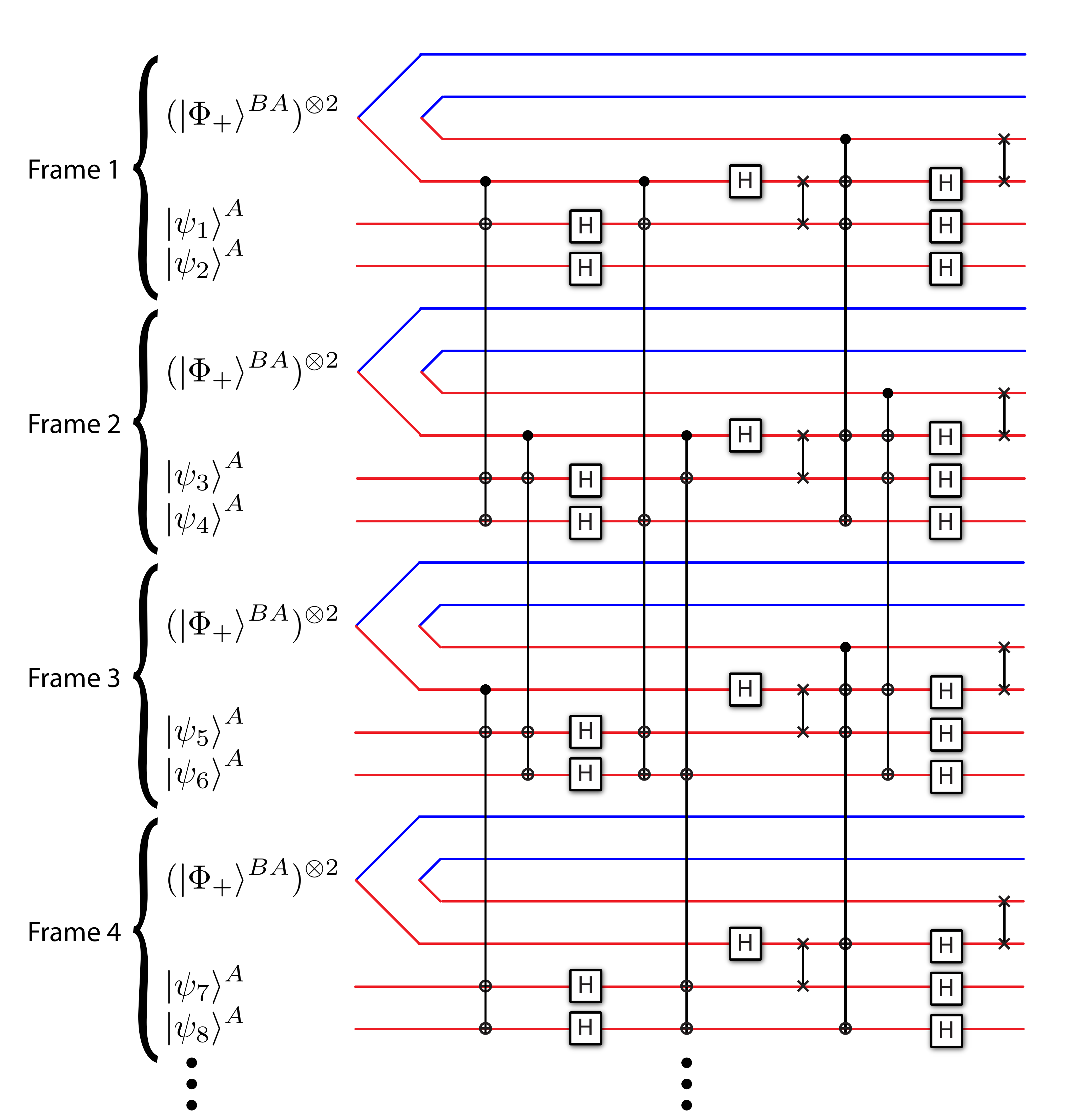}%
\caption{(Color online) An online encoding circuit for an
entanglement-assisted quantum convolutional code. The receiver Bob possesses
the first two qubits in the two ebits and the sender Alice possesses the
second two qubits in the two ebits. The sender encodes two information qubits
per frame with the help of her half of the two ebits.}%
\label{fig:science-conv}%
\end{center}
\end{figure}
Figure~\ref{fig:science-conv}\ illustrates the operations that transform the
unencoded stabilizer to the encoded one in an online encoding circuit (See
Refs.~\cite{isit2006grassl,arx2007wildeEAQCC} for details of translating
gate-level operations to polynomial operations on the stabilizer matrix). The
final stabilizer is as follows:%
\[
\left[  \left.
\begin{array}
[c]{cccccc}%
0 & 1 & 0 & 1 & 0 & 0\\
1 & 0 & 0 & 0 & 1 & 0\\
0 & D & l\left(  D\right)  & D & 1 & D\\
f\left(  D\right)  & h\left(  D\right)  & 0 & 1 & 0 & 0
\end{array}
\right\vert
\begin{array}
[c]{cccccc}%
0 & 0 & 0 & 0 & 0 & 0\\
0 & 0 & 0 & 0 & 0 & 0\\
0 & 1 & 0 & 1 & 0 & 0\\
1 & l\left(  D\right)  & l\left(  D\right)  & l\left(  D\right)  & 1 & D
\end{array}
\right]  ,
\]
where $h\left(  D\right)  \equiv D^{-1}+1+D$ and $l\left(  D\right)
\equiv1+D$. Compare Alice's generators in the last two rows of the above
matrix to the quantum check matrix in (\ref{eq:multigen-stab}). We have
constructed a code with the same error-correcting properties because these two
matrices are equivalent. The above code is a $\left[  \left[  4,2;2\right]
\right]  $ entanglement-assisted convolutional code because it encodes two
information qubits and consumes two ebits for every four uses of the noisy
quantum channel.

It is interesting to compare the above example to the similar example in
Section VII of Ref.~\cite{arx2008wildeGEAQCC}. In that example, we expanded
the generators by a factor of two and were able to encode six information
qubits and consume two ebits for every eight uses of the channel. The code in
Ref.~\cite{arx2008wildeGEAQCC} makes a more efficient use of entanglement at
the price of a doubling of the frame size (and thus a doubling of the memory
requirements). In addition, the code in Ref.~\cite{arx2008wildeGEAQCC}
requires infinite-depth operations, but the code in the above example uses
finite-depth operations only.

\section{Classical Enhancement}

It is possible to \textquotedblleft piggyback\textquotedblright\ classical
information along with the extra entanglement in a fashion similar to the
superdense coding effect \cite{PhysRevLett.69.2881}. One can then decide to
send classical information only or combine the transmitted classical
information with extra entanglement to teleport more qubits
\cite{PhysRevLett.70.1895}. We mention that
Refs.~\cite{Kremsky:2008:012341,arx2008wildeUQCC} offer a different way of
piggybacking classical information along with an entanglement-assisted code.

Consider the stabilizer in (\ref{eq:final-encoded-Z}-\ref{eq:final-encoded-X}%
). Recall that the first two sets of rows correspond to \textquotedblleft
extra entanglement\textquotedblright\ generators and the second two sets of
rows are their corresponding generators. It is only clear to us that this
piggybacking effect can work for rows that are not affected by infinite-depth
operations---the first and third sets of rows in (\ref{eq:final-encoded-Z}%
-\ref{eq:final-encoded-X}).

Consider the first and third sets of rows in (\ref{eq:final-encoded-Z}%
-\ref{eq:final-encoded-X}):%
\begin{equation}
\left[  \left.
\begin{array}
[c]{ccc}%
I & 0 & I\\
E_{1,a}^{\prime\prime}\left(  D\right)  & 0 & E_{1,a}^{\prime}\left(
D\right)
\end{array}
\right\vert
\begin{array}
[c]{ccc}%
0 & 0 & 0\\
\Gamma_{1}\left(  D\right)  & 0 & \Gamma_{1}\left(  D\right)
\end{array}
\right]  , \label{eq:classical-enhancement}%
\end{equation}
where we omit the last four sets of columns of zeros and label the respective
submatrices of $E_{1}^{\prime\prime}\left(  D\right)  $ and $E_{1}^{\prime
}\left(  D\right)  $ as $E_{1,a}^{\prime\prime}\left(  D\right)  $ and
$E_{1,a}^{\prime}\left(  D\right)  $. Consider the following set of operators:%
\begin{equation}
\left[  \left.
\begin{array}
[c]{ccc}%
0 & 0 & E_{1,a}^{\prime T}\left(  D^{-1}\right)  \Gamma_{1}\left(  D\right)
\end{array}
\right\vert
\begin{array}
[c]{ccc}%
0 & 0 & I
\end{array}
\right]  . \label{eq:classical-ops}%
\end{equation}
The rows in (\ref{eq:classical-ops}) anticommute with the first set of rows in
(\ref{eq:classical-enhancement}) but commute with the second set of rows in
(\ref{eq:classical-enhancement}) according to the shift symplectic product
\cite{arx2007wilde,arx2007wildeEAQCC}. Furthermore, the rows in
(\ref{eq:classical-ops}) anticommute with the first set of rows in
(\ref{eq:classical-enhancement}) in such a way that the result of the shifted
symplectic product is the identity matrix because%
\[
I\cdot I+0\cdot\Gamma_{1}^{T}\left(  D^{-1}\right)  E_{1,a}^{\prime}\left(
D\right)  =I.
\]
The operators in (\ref{eq:classical-ops}) therefore give Alice a way to send
$s$ extra classical bits in analogy to superdense coding (we define $s$ at the
beginning of the proof of the main theorem). Alice sets any of the $s$ bits to
\textquotedblleft1\textquotedblright\ by applying the corresponding
\textquotedblleft encoded\textquotedblright\ version of the operator in
(\ref{eq:classical-ops}) (one can determine the encoded version of the
operator by applying the rest of the column operations of the encoding circuit
to the operator). The fact that the generators in (\ref{eq:classical-ops})
commute with the second set of rows in (\ref{eq:classical-enhancement})
guarantees that encoding this extra classical information does not
\textquotedblleft throw off\textquotedblright\ the operation of the code for
error correction. Bob simply has to measure the encoded version of the
operators corresponding to the first set of rows in
(\ref{eq:classical-enhancement}) to determine the values of the $s$\ classical bits.

One can use this classical enhancement just to send classical information and
the result is an $\left[  \left[  n,k:s;c\right]  \right]  $
classically-enhanced entanglement-assisted convolutional code according to the
notation of Ref.~\cite{Kremsky:2008:012341}. Alternatively, one can use the
classical communication and extra entanglement to teleport more quantum
information \cite{PhysRevLett.70.1895}. Teleporting then gives an $\left[
\left[  n,k+s/2;c+s/2\right]  \right]  $ entanglement-assisted convolutional code.

We mention that there is no \textquotedblleft free lunch\textquotedblright%
\ with this classical enhancement technique. Exploiting the technique reduces
number of errors that the code corrects. The encoded versions of the operators
in (\ref{eq:classical-ops}) are actually a basis for the extra errors that the
code corrects. Bob measures the encoded version of the operators corresponding
to the first set of rows in (\ref{eq:classical-enhancement}) to retrieve the
syndrome bits for these errors. It may be useful to correct these additional
errors, but typically, one begins with a given set of generators that have
desirable error-correcting properties. If one chooses to construct the
entanglement-assisted quantum code with the techniques developed in the
previous section, the resulting quantum code possesses the original desired
error-correcting properties. One can then choose whether to exploit the extra
entanglement for extra error-correcting capability or for classical enhancement.

\subsection{Classical Enhancement of the Example}

We show how to enhance the example so that it also sends classical
information. Consider the following two operators:%
\[
\left[  \left.
\begin{array}
[c]{cccccc}%
0 & 0 & D^{-1} & 1+D+D^{-2} & 0 & 0\\
0 & 0 & 0 & 1+D^{2} & 0 & 0
\end{array}
\right\vert
\begin{array}
[c]{cccccc}%
0 & 0 & 1 & 0 & 0 & 0\\
0 & 0 & 0 & 1 & 0 & 0
\end{array}
\right]  .
\]
The first row anticommutes with the first row in
(\ref{eq:unencoded-free-ent-example}) and commutes with all other rows in
(\ref{eq:unencoded-free-ent-example}). The second row anticommutes with the
second row in (\ref{eq:unencoded-free-ent-example}) and commutes with all
other rows in (\ref{eq:unencoded-free-ent-example}). These commutation
relations imply that the above operators are useful for encoding classical
information in a superdense-coding-like fashion. These operators encode two
classical bits into the code and make use of the first two rows in
(\ref{eq:unencoded-free-ent-example}) instead of just \textquotedblleft
wasting\textquotedblright\ them. Measuring the first two rows in
(\ref{eq:unencoded-free-ent-example}) reveals the values of the two classical
bits. We can determine the encoded versions of these \textquotedblleft
classical-information-encoding\textquotedblright\ operators by tracing how the
operators change in the Heisenberg picture through the rest of the encoding
circuit. The result is a $\left[  \left[  4,2:2;2\right]  \right]  $
classically-enhanced entanglement-assisted convolutional code.

Alternatively, we can use these two classical bits and consume one ebit to
teleport an additional information qubit. This technique produces a $\left[
\left[  4,3;3\right]  \right]  $ entanglement-assisted convolutional code.

\section{Conclusion}

We have constructed a theory of entanglement-assisted quantum convolutional
coding for codes that do not have the CSS\ structure. The method of this paper
uses entanglement less efficiently than the protocol in
Ref.~\cite{arx2008wildeGEAQCC}, but it does not require expanding a set of
generators and therefore does not require a heuristic convergence argument as
do the codes in Ref.~\cite{arx2008wildeGEAQCC}. The \textquotedblleft extra
entanglement\textquotedblright\ method results in encoding and decoding
circuits that act on smaller numbers of qubits than the circuits in
Ref.~\cite{arx2008wildeGEAQCC} and therefore requires less memory and circuit
complexity. It should be of interest to find solutions in between the
entanglement-efficient codes \cite{arx2008wildeGEAQCC} and the
entanglement-inefficient codes discussed in this paper.

It would be ideal to know the exact trade-off between entanglement for the
codes in this paper and the entanglement for the codes in
Ref.~\cite{arx2008wildeGEAQCC}, but we cannot establish this relationship
right now because it is still an open question to determine the exact amount
of entanglement that the codes in Ref.~\cite{arx2008wildeGEAQCC} require.
Also, it would be ideal to know the trade-off between the frame size for codes
in this paper and the frame size for codes in Ref.~\cite{arx2008wildeGEAQCC}.
Note that it also remains an open question to determine the frame size for codes in
Ref.~\cite{arx2008wildeGEAQCC} because we have not yet shown the exact step at
which the algorithm from Ref.~\cite{arx2008wildeGEAQCC} converges.

It may be possible to avoid either using extra entanglement as outlined in
this paper or expanding the generator set as outlined in
Ref.~\cite{arx2008wildeGEAQCC}; we have no proof that these constructions are
optimal. The fact that CSS codes do not need these methods provides some
evidence that it may be possible \cite{arx2007wildeEAQCC}. However, after
extensive exploration we have not found such a better technique, making us
believe that it is unlikely.

The authors thank Hari Krovi, Markus Grassl, and Martin R\"{o}tteler for
stimulating discussions. MMW\ acknowledges support from NSF Grant 0545845 and
the National Research Foundation \& Ministry of Education, Singapore,\ and
TAB\ acknowledges support from NSF Grant CCF-0448658 and NSF Grant ECS-0507270.

\bibliographystyle{apsrev}
\bibliography{qccfe}

\end{document}